\documentclass[preprint,showpacs,preprintnumbers,amsmath,amssymb]{revtex4}
\usepackage{graphicx,amsmath,amssymb,amsfonts,latexsym,color,dcolumn,bm,epsfig,subfigure} 
\newcommand{\be}{\begin{equation}}
\newcommand{\ee}{\end{equation}}
\newcommand{\bi}{\begin{itemize}}
\newcommand{\ei}{\end{itemize}}
\newcommand{\eenn}{\nonumber \end{equation}}
\newcommand{\beqn}{\begin{eqnarray}}
\newcommand{\eeqn}{\end{eqnarray}}

\def\braket#1{\mathinner{\langle{#1}\rangle}}

{\catcode`\|=\active\gdef\Braket#1{\left<\mathcode`\|"8000\let|\bravert {#1}\right>}} 
\def\bravert{\egroup\,\vrule\,\bgroup}
\begin{document}

\title{High-momentum tail in the Tonks-Girardeau gas under general confining potentials}

\author{Gustavo A. Moreno}
\affiliation{CONICET and Departamento de F\'{i}sica, FCEN, Universidad de Buenos Aires, Ciudad
Universitaria, 1428 Buenos Aires, Argentina}

\begin{abstract}
We prove that the ground state momentum distribution of a one-dimensional system of impenetrable bosons exhibits a $k^{-4}$ tail for any confining potential. We also derive an expression for easily computing the asymptotic occupation numbers and verify our results with an exact numerical approach.
\end{abstract}

\date{\today}

\maketitle

\section{Introduction}
It is well known, since Girardeau original paper \cite{girardeau}, that one-dimensional impenetrable point-like bosons can be mapped into a free fermion problem taking the product of the many-body wave function of the fermion system with a unit antisymmetric function. Although restricted to one dimension, this remarkable observation opened the possibility of treating analytically a strongly interacting problem. As can be easily seen local observables (such as the probability of finding the particles in a well defined position) remain the same in both systems, but non-local quantities of the bosonic problem need to be calculated in a non-trivial way. In this letter we focus on the one-body reduced density matrix $n(x,y)$ for an $N$-body system of bosons at $T=0$:
\be\label{eq:definition}
n(x,y)=N \int \psi^*(x,x_2, \ldots, x_N) \,\psi(y,x_2, \ldots, x_N) \,dx_2, \ldots, dx_N ,
\ee 
where $\psi$ is the many-body wave function for the ground state (unit norm). As pointed out in \cite{girardeau} it is not an easy task to reveal the properties of this object because the multiple integration can not be carried out explicitly. Difficulties remain in quantities such as the momentum distribution (MD), which can be found projecting equation (\ref{eq:definition}) onto a plane wave basis. The low momentum region of the MD was explored by Schultz and Lenard \cite{schultz,lenard} to determine whether Bose-Einstein condensation occur in the thermodynamic limit, as argued by Girardeau \cite{girardeau}. However, at that time, the interest in the model was mainly theoretical and two decades pass until further analytical results concerning the asymptotic properties of the momentum distribution were published \cite{vaidya}.\\
This situation changed ten years after the first successful experiments on Bose-Einstein condensation of alkali gases \cite{becs}. These last provided the framework for achieving effectively one-dimensional ($T\approx 0$) bosonic systems \cite{olshanii0} in laboratory experiments \cite{paredes, kinoshita}. After its experimental realization, further properties of the momentum distribution were extensively studied (mainly numerically) not only for theoretical reasons but because the MD characteristic features helped to determine whether the system had entered into the strongly interacting regime.\\
In the following we show an asymptotic expansion for the one-body reduced density matrix valid for short-range correlations and we derive an universal power law for the tail of the momentum distribution, namely $k^{-4}$. This behaviour was found previously only for particular potentials \cite{lapeyre,minguzzi,forrester} or constrained to formulations on the lattice \cite{rigol}.\\
The goal of this paper is to give a general and self-contained proof of this power law valid for any confining potential. In addition, we derive an expression which allows an easy computation of the short distance properties of the one-body reduced density matrix. \\

\section{Short distance correlations}
\begin{figure}[htb]
\centerline{\epsfig{file=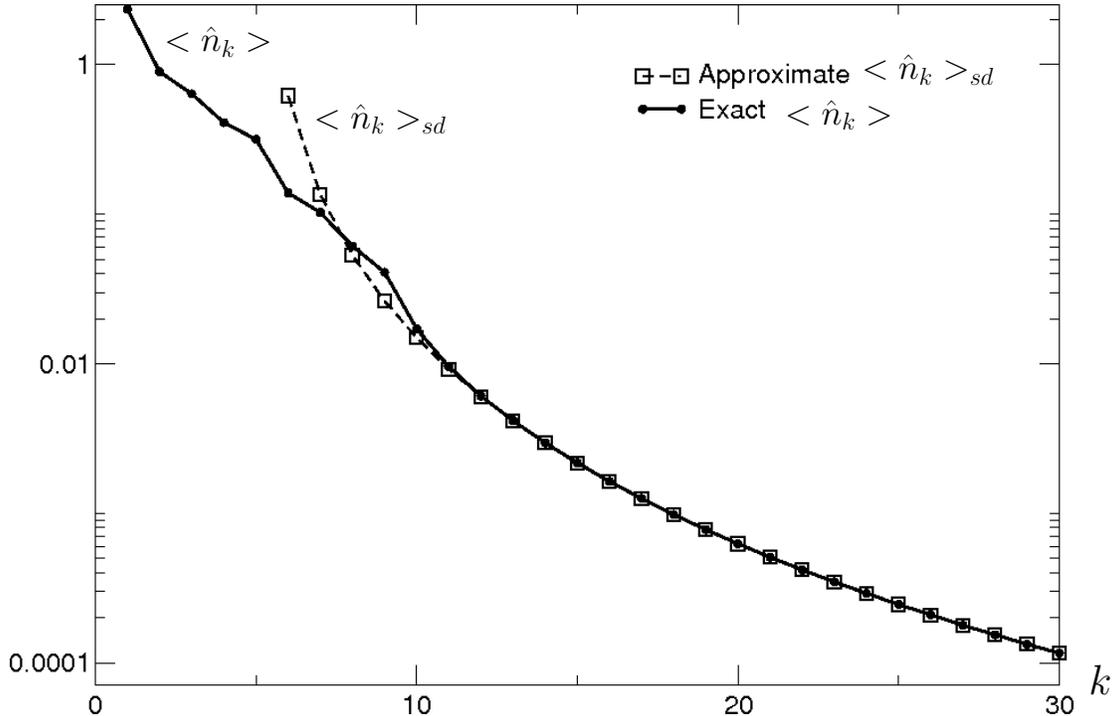,width=0.9\textwidth,angle=0}}
\caption{Numerical evaluation of the exact distribution for $5$ bosons in a flat box $\braket{\hat n_k}$. Comparison with the approximated short-distance expansion $\braket{\hat n_k}_{sd}$ (from equation (\ref{eq:analytic})). Note that both axes are dimensionless.}
\label{fig:fig1} 
\end{figure}

Consider a one-dimensional system of $N$ impenetrable point-like bosons trapped by a potential $V(x)$. As shown by Girardeau \cite{girardeau} the ground state many-body wave function can be constructed as:
\be
\psi(x_1, \ldots, x_N)=\frac{1}{\sqrt{N!}}\,A(x_1,\ldots,x_N)\,\det{\left[\phi_i(x_j)\right]}, 
\ee
where $A(x_1,\ldots,x_N)$ is the unit antisymmetric function and $\{\phi_i\}_{i=1}^{N}$ are the $N$ lowest eigenstates of the one-body Hamiltonian:
\be
E_i \,\phi_i(x)= -\frac{\hbar^2}{2m} \frac{d^2\phi_i}{dx^2} \,+V(x)\,\phi_i(x),
\ee
which can be taken as real functions. To simplify our notation we will continue to consider the case of the ground state but the same arguments will hold, with minor modifications, for any other pure state of the system (see Section \ref{sec:tail}).\\
Definition (\ref{eq:definition}) can be written as:
\beqn 
&& n(x,y)=\frac{1}{(N-1)!}\int sgn(x-x_2) \ldots sgn(x-x_N)\,sgn(y-x_2)\ldots\times \nonumber \\ && \ldots sgn(y-x_N) \left(\sum_{\sigma \in \mathbb{S}_N} Sgn(\sigma) \,\phi_{\sigma(1)}(x)\, \phi_{\sigma(2)}(x_2)\ldots \phi_{\sigma(N)}(x_N)\right) \times \nonumber \\
&& \left(\sum_{\lambda \in \mathbb{S}_N} Sgn(\lambda)\,\phi_{\lambda(1)}(y)\, \phi_{\lambda(2)}(x_2)\ldots \phi_{\sigma(N)}(x_N)\right) \,\,dx_2, \ldots, dx_N
\,\, ,\eeqn
where $sgn(x)$ stands for the unit antisymmetric function and $Sgn(\sigma)$ for the sign of the permutation $\sigma \in \mathbb{S}_N$. This equation can be rearranged in the following form (assuming, with no loss of generality, that $y \geq x$):
\beqn \label{eq:exact}
n(x,y)=\frac{1}{(N-1)!}\sum_{\sigma,\lambda \in \mathbb{S}_N}\phi_{\sigma(1)}(x)\phi_{\lambda(1)}(y)\times \nonumber \\\,Sgn(\sigma)\,Sgn(\lambda)\,
\prod_{j=2}^N\left(\delta_{\sigma(j),\lambda(j)} \,-2\int_{x}^{y}\phi_{\sigma(j)}(z)\phi_{\lambda(j)}(z)\,dz\right) \,.
\eeqn
Equivalently one can write an expression removing the constraint $y \geq x$, namely:
\beqn \label{eq:exact2}
n(x,y)=\frac{1}{(N-1)!}\sum_{\sigma,\lambda \in \mathbb{S}_N}\phi_{\sigma(1)}(x)\phi_{\lambda(1)}(y)\,Sgn(\sigma)\,Sgn(\lambda)\,\times \nonumber \\
\prod_{j=2}^N\left(\delta_{\sigma(j),\lambda(j)} \,-2[\theta(y-x)-\theta(x-y)]\int_{x}^{y}\phi_{\sigma(j)}(z)\phi_{\lambda(j)}(z)\,dz\right) \, ,
\eeqn
where $\theta(x)$ is the Heaviside step function. This expression is exact an can be used to calculate $n(x,y)$ numerically when the eigenfunctions are known (or calculable). Note that the product of all Kronecker-delta terms corresponds to a free fermion distribution and the remaining ``bosonic correction'' appears with powers of $\int_x^y$. To proceed further we note that short distance properties of this expression should be well approximated to first order in $\int_{x}^{y}$, because the projection of large scales will be strongly suppressed in the high momentum regime (observation stated without proof in \cite{pezer}). This is equivalent to saying that occupation numbers on a basis $\{\chi_i\}_{i\geq1}$ (not necessarily an eigenbasis of the Hamiltonian), $\braket{\hat n_k}=\int\chi_k^*(x)n(x,y)\chi_k(y) dxdy$, will be well approximated by the first order term of $n(x,y)$ if $k>>N$ ($N$ fixes a maximum energy scale in the system). As we shall see in the last section these facts are correct, the leading correction is of first order in $\int_x^y$ and the remaining terms with more than one integral can be neglected. We will first explore the consequences of this argument and leave the details of the proof to the last section.\\
As claimed before, performing the sum up to first order in $\int_{x}^{y}$ yields an approximate expression ($n_{sd}$) valid for describing the short distance properties of the system ($y \geq x$):
\beqn \label{eq:main}
&& n_{sd}(x,y)= \sum_{i=1}^N\phi_i(x)\phi_i(y) - 2\sum_{i,j=1}^N\int_{x}^{y}\phi_i(x)\phi_i(y)\phi_j(z)\phi_j(z)dz + \nonumber \\ &&
+ 2\sum_{i,j=1}^N\int_{x}^{y}\phi_i(x)\phi_j(y)\phi_i(z)\phi_j(z)dz \,\,\,\, . 
\eeqn
Note that no permutations are involved in this sum, which makes it easier to handle. As an example of the applicability of this expression to study short distance properties let us compute the MD for a flat confining potential, which has been extensively studied in literature. The approximate expression for the occupation numbers in a box of unit length can be evaluated as: $\braket{\hat n_k}_{sd}=\int \chi_k^*(x)n_{sd}(x,y)\chi_k(y)\,dxdy$, where $\chi_k(x)=\{\sqrt{2}\sin(\pi k x)\}_{k\geq1}$. This expression involves simple combinations of trigonometric functions and can be evaluated explicitly. We have carried out this calculation finding:
\be \label{eq:analytic}
\braket{\hat n_k}_{sd}=\frac{4}{\pi^2}\sum_{i,j=1, \,i\neq j}^{N} \frac{2k^2i^2-k^2j^2-i^2j^2}{(k^2-i^2)^2(k^2-j^2)}.
\ee
As a first check of our approximation we have taken the thermodynamic limit of equation (\ref{eq:analytic}), finding $\braket{\hat n_k}\approx \frac{4}{3\pi^2}(\frac{N}{k})^4$. This expression agrees with previous calculations after the identification of $N$ with the Fermi level $k_F$\cite{forrester}.\\
The convergence of the approximated MD to its exact value is rather fast. We have verified this property comparing the results of the approximation with the exact ones (equation (\ref{eq:exact}) or (\ref{eq:exact2})) for a finite number of particles. As a concrete example we computed the exact momentum distribution for a system of $5$ bosons in the flat one-dimensional box. We exhibit these results in Figure \ref{fig:fig1} where we plot the exact numerical evaluation $\braket{\hat n_k}$ and the approximate one $\braket{\hat n_k}_{sd}$. As argued before, when $k>>N$, the agreement between both expressions is excellent.\\

Finally we note that equation (\ref{eq:analytic}) can be approximated for large $k$ by an expression of the form $k^{-4}+\mathcal{O}(k^{-6})$. In the next section we show that this is a general characteristic of a confined Tonks-Girardeau (TG) gas.  

\section{Asymptotic tail}\label{sec:tail}
Let us now show that the momentum distribution decays as $k^{-4}$ regardless of the confining potential. For a confined system of $N$ hard-core bosons in a general potential $V(x)$ take $\{\phi_i\}_{i=1}^{N}$ to be the set of the $N$ lowest energy eigenfunctions. As argued previously the occupation numbers for large momentum $k$ can be determined projecting equation (\ref{eq:main}) for $n_{sd}$ onto a plane wave basis. The projection of the first term in that equation can be disregarded in almost all situations. In fact, if the potential is a smooth function, say $\mathcal{C}^\alpha$ piecewise, then any eigenfunction is at least $\mathcal{C}^{\alpha+2}$. This means that the coefficients of the projection will be of order $1/k^{2(\alpha+2)}$. So, for smooth functions $V(x)$ the asymptotic behaviour is determined by the second and third terms. From now on we omit the first term in equation (\ref{eq:main}). Then, coming back to:
\be\label{eq:occupation}
\braket{\hat n_k}_{sd}=\int \chi_k^*(x)n_{sd}(x,y)\chi_k(y)\,dxdy 	\,,
\ee
we explicitly remove the condition $x<y$ from the expressions. In order to do this we write:
\be
n_{sd}(x,y)=\mathbb{N}(x,y)[\theta(y-x)-\theta(x-y)] \,\,\,\,\, ,
\ee
where $\forall\, x,\,y$
\beqn
\mathbb{N}(x,y)=- 2\sum_{i,j}\int_{x}^{y}\phi_i(x)\phi_i(y)\phi_j(z)\phi_j(z)dz \,+\nonumber\\+ 2\sum_{i,j}\int_{x}^{y}\phi_i(x)\phi_j(y)\phi_i(z)\phi_j(z)dz  \,\, .
\eeqn
To perform the integral (\ref{eq:occupation}) in a plane wave basis $\chi_k=\frac{e^{ikx}}{\sqrt{L}}$ we change variables to:
\be
\left\{ \begin{array}{c}
x=X \\ y=X+r .
\end{array}\right.
\ee
Now we have:
\beqn
&& \braket{\hat n_k}_{sd}=\frac {1}{L}\int e^{ikr} \,\mathbb{N}(X,r)\,[\theta(r)-\theta(-r)]\,dXdr= \nonumber \\ && =\int e^{ikr} \left(\frac {1}{L}\int\,\mathbb{N}(X,r)dX\right)\,[\theta(r)-\theta(-r)]\,dr\, . 
\eeqn
The power law for the high momentum regime can be found expanding $\mathbb{N}(X,r)$ in Taylor series around $r=0$. Surprisingly enough a direct calculation shows that the function $\mathbb{N}(X,r)$ verifies:
\be \label{eq:conditions}
\left\{ \begin{array}{c} 
\mathbb{N}(X,0)=0 \\ \\ \frac{\partial \mathbb{N}}{\partial r}(X,0)=0 \\ \\ \frac{\partial^2 \mathbb{N}}{\partial r^2}(X,0)=0 \\ \\ \frac{\partial^3 \mathbb{N}}{\partial r^3}(X,0)\neq0
\end{array}\right.    .
\ee
Which is a consequence of its symmetries. Taking (\ref{eq:conditions}) into account, one ends with (up to leading order in $r$):
\be
\braket{\hat n_k}_{sd}\propto \int e^{ikr} r^3\,[\theta(r)-\theta(-r)]dr=\frac{1}{i^3}\frac{\partial^3}{\partial k^3} \int e^{ikr}[\theta(r)-\theta(-r)]dr .
\ee
Which is equivalent to $\braket{\hat n_k}_{sd}\propto k^{-4}$.\\
Notice that the same argument holds when excited states of the gas are considered. In such situation the set of energy eigenfunctions of the one-body problem which are used to calculate the many-body wave function will have one element which corresponds to the largest eigenenergy. This element fixes the scale $k_0$ (maximum typical wave number involved in that energy scale) such that equation (\ref{eq:main}) becomes a good approximation in the region $k\gg k_0$ (in the numerical example of the previous section $k_0=N$).
\section{Technical details}
In the previous section it was shown that the leading order contribution of the term with one integral is of order $r^3$. As a consequence of this fact the approximate expression (7) would be valid only if we prove that the terms with two and three integrals vanish in equation (\ref{eq:exact2}), which is a non-trivial fact (note that those terms can, in principle, contribute to order $r^3$). To see this, we first note that the terms with two integrals have a prefactor $[\theta(y-x)-\theta(x-y)]^2=1$, which is non-singular and thus can be neglected as we argued for the fermion term in equation (9). Secondly, for terms with three integrals one gets:
\beqn
\sum_{\sigma,\lambda \in \mathbb{S}_N}\phi_{\sigma(1)}(x)\phi_{\lambda(1)}(y)\,[\theta(y-x)-\theta(x-y)]\, Sgn(\sigma)\,Sgn(\lambda)\,\times \nonumber \\ 
\left(\int_{x}^{y}\phi_{\sigma(a)}(z)\phi_{\lambda(a)}(z)\,dz\right)\,\left(\int_{x}^{y}\phi_{\sigma(b)}(z)\phi_{\lambda(b)}(z)\,dz\right)\,\times \nonumber \\ 
\left(\int_{x}^{y}\phi_{\sigma(c)}(z)\phi_{\lambda(c)}(z)\,dz\right) \prod_{j=2, j\neq a,b,c }^N \delta_{\sigma(j),\lambda(j)} \, , 
\eeqn
where the complete term is the sum over all choices of the integers $a$, $b$ and $c$. For each fixed choice the contribution to the order $r^3$ is found by replacing each integral to first order in $r$. The resulting prefactor reads:
\beqn
\sum_{\sigma,\lambda \in \mathbb{S}_N}Sgn(\sigma)\,Sgn(\lambda)\,\phi_{\sigma(1)}(X)\phi_{\lambda(1)}(X)\times \nonumber \\ \phi_{\sigma(a)}(X)\phi_{\lambda(a)}(X)
\phi_{\sigma(b)}(X)\phi_{\lambda(b)}(X)\phi_{\sigma(c)}(X)\phi_{\lambda(c)}(X) \,\, ,
\eeqn
which vanishes when summing over all permutations. This completes the proof.\\

\section{Concluding Remarks}
We have shown that the momentum distribution of a TG gas in the high momentum limit can be described by a power law of the form $k^{-4}$ regardless of the confining potential. The insight given by equation (\ref{eq:main}) shows that this universal behaviour is a consequence of the wave function symmetries together with the cusp condition imposed by the subsidiary requirement $\psi(x_1,\ldots, x_n)=0$ if $x_i=x_j$. \\
We have also proven the validity of equation (\ref{eq:main}), which is found to be a suitable alternative for an accurate and numerically-efficient description of the high-momentum region of the TG gas.\\

\section{Acknowledgments}
I am grateful to P. L. De N\'apoli and E. Calzetta for useful discussions and to A. Caso and J. Peralta Ramos for valuable comments. 
I acknowledge the financial support of CONICET (Argentina). 


\end{document}